\documentclass[useAMS,usegraphicx,usenatbib]{mn2e} 
\usepackage{color,ulem}

\newcommand{\eqb}{\begin{eqnarray}}
\newcommand{\eqe}{\end{eqnarray}}
\newcommand{\eqen}{\nonumber\end{eqnarray}}

\def\b2{B2 0902+34}

\def\gpeak{\gamma_{\rm peak}}
\def\npeak{N_{\rm peak}}
\def\numax{\nu_{\rm max}}
\def\lmax{L_{\nu_{\rm max}}}
\def\bequip{B_{\rm equip}}

\begin{document}

\title[Energies from curved synchrotron spectra]
{The non-thermal emission of extended radio galaxy lobes with curved electron spectra}
\author[Duffy \& Blundell]
{\parbox[]{6.in} {Peter Duffy$^{1}$ and Katherine M.\ Blundell$^{2}$ }\\\\ 
\footnotesize
$^{1}${UCD School of Physics, University College Dublin,
  Dublin 4, Ireland}\\
$^{2}${University of Oxford, Astrophysics, Keble
  Road, Oxford, OX1 3RH, U.K.}}
\maketitle

\begin{abstract}
  The existing theoretical framework for the energies stored in the
  synchrotron-emitting lobes of radio galaxies and quasars doesn't
  properly account for the curved spectral shape that many of them
  exhibit.  We characterise these spectra using parameters that are
  straightforwardly observable in the era of high-resolution,
  low-frequency radio astronomy: the spectral curvature and the
  turnover in the frequency spectrum.  This characterisation gives the
  Lorentz factor at the turnover in the energy distribution (we point out that this is distinctly
  different from the Lorentz factor corresponding to the turnover
  frequency in a way that depends on the amount of curvature in the
  spectrum) and readily gives the equipartition magnetic field
  strength and the total energy of the radiating plasma obviating the
  need for any assumed values of the cutoff frequencies to calculate
  these important physical quantities.  This framework readily yields
  the form of the X-ray emission due to inverse-Compton (IC)
  scattering of Cosmic Microwave Background (CMB) photons by the
  electrons in the plasma having Lorentz factors of $\sim$1000.  We
  also present the contribution to CMB anisotropies due to
  relativistic plasmas such as giant radio galaxy lobes, expressed in
  terms of the extent to which the lobes have their magnetic field and
  particle energies are in equipartition with one another.
\end{abstract}

\section{Introduction}
\label{sec:intro}

In order to quantify the feedback and influence of the jets and lobes of
radio galaxies and quasars in terms of the energy and heat they inject
into their environments (hence their influence on cosmic structure
formation) and also their contamination to primordial anisotropies in
the Cosmic Microwave Background (CMB) it is necessary to understand
the nature of the underlying particle energy spectrum of these
plasmas.  For example, to calculate the energy stored in radio galaxy
lobes, it is necessary to integrate over the entire energy
distribution of the particles in the plasma which constitutes the
lobes.  Traditionally, this is calculated by assuming that these
plasmas have very simple power-law energy distributions \citep[e.g.\
][]{Long81,Miley80} but there are two distinct drawbacks of this
approach: the first problem is that radio spectra are frequently
observed to be curved rather than power-law \citep[e.g.\
][]{Laing1980,Landau86,Car91,Blu99,Klamer2006,Jamrozy2008}.  The
second problem is that the lower limit of  the power-law
approximation of this distribution of energies (expressed in terms of
Lorentz factor $\gamma$), also known as the low-energy cutoff or
turnover, referred to as $\gamma_{\rm min}$ has been unconstrained for
decades (e.g.\ $\gamma_{\rm min}$ = 1 is used by \citet{Hard2005} and
also by \citet{Kaiser1997}, $\gamma_{\rm min}$ = 10 is used by
\citet{Cro05}, $\gamma_{\rm min}$ = 100 is used by \citet{Car91}, 1000
is taken by \citet{War98}); uncertainty on this scale of the
low-energy cutoff leads to orders of magnitude uncertainty in the
estimate of the energy stored by the plasma reservoirs 
\cite[][]{Mocz11}.

We present a new formalism which overcomes the difficulties previously
encountered in inferring energies and magnetic fields from
synchrotron-emitting plasma exhibiting curved spectra, characterising
this curved spectral shape with curvature coefficients that are
straightforwardly fitted in terms of the observed peak frequency
($\nu_{\rm peak}$) and the observed curvature ($q$) giving simple
expressions for the magnetic field strength $(B)$ and the total energy
density ($e)$. We demonstrate consistency with the traditional
power-law case in the appropriate limit. We present a specific
discussion of how the magnetic field itself can be determined
 if equipartition is assumed and quantify the extent to which
these plasma lobes are contaminants of primordial CMB anisotropies on
the relevant angular scales, noting that this extent is a strong
function of whether the plasma lobes have their magnetic field and
particle energies in equipartition.

\section{The Curved Spectra of Classical Double Radio Sources}

\citet{Blu99} found for the complete samples of low-frequency
selected classical double radio galaxies they
analysed across a wide range of frequencies that $\sim 70$\% of these
objects have curved synchrotron spectra at radio wavelengths.  They
fitted these spectra of frequency $\nu$ and emissivity $L_{\rm \nu}$
to functions of the following form:

\begin{equation}
\ln(L_\nu)=a_0+a_1\ln(\nu)+a_2[\ln(\nu)]^{2}.
\label{eq:curvedspec_lum_nu}
\end{equation}

The consistently curving spectra of such sources have historically
been neglected when their energies and magnetic field strengths have
been estimated via the \citet{Long81} and \citet{Miley80} formalism.
Often it has been assumed that curvature is simply a high-energy
phenomenon only arising from synchrotron ageing in the GHz regime.
However, there are no definitive examples of a power-law spectrum
being observed at low radio frequencies with bolt-on curvature (such
as that from synchrotron ageing) only being observed at higher radio
frequencies.  Indeed, \citet{Blu00} explained why the origin of a
curved spectrum cannot be attributed to the ageing of a
synchrotron spectrum and so the paradigm of an 
intrinsically curved electron-energy
spectrum $N(\gamma)$ cannot be dismissed.  Studies of complete samples
of classical double radio sources showed that the dominant energy-loss
mechanism for radio lobes are not synchrotron losses but adiabatic
expansion losses \citep{Blu99} which are energy-independent
\citep{Sch68}.  It is most interesting in this regard that
\citet{Rudnick1994}, in a detailed analysis of the spatially resolved
spectral shape of the famous classical double radio galaxy
Cygnus A, found {\it no evidence for any variation in the curvature}
of $N(\gamma)$ throughout even the oldest or youngest regions of this
prototypical object.  This points to an intrinsically curved particle spectrum
in the plasma lobes of Cygnus\,A.

We briefly review and reject three possibilities of this spectral
curvature being caused by external factors, rather than arising from
the inherent particle distribution: (i) Free-free absorption: in
principle, free-free absorption of synchrotron photons by the medium
surrounding the radio lobes could cause spectral curvature. However,
since many radio galaxy and quasar lobes have lengths of 100s of kpc
\citep[e.g.\ ][]{Blu99} which lie way beyond their host galaxies, this
explanation can be dismissed as a dominant effect except for the
innermost 10s of kpc of which might have sufficient gas densities for
this absorption to be significant. (ii) Synchrotron self-absorption:
if the plasma is optically thick to synchrotron photons, as may be the
case at very low radio frequencies, then this could dramatically curve
the spectra. However, this would produce rising spectra not falling
spectra characterised by the sources whose spectra we focus on in this
paper.  (iii) The Razin effect might in principle choke off
synchrotron radiation, if the plasma particle density were
sufficiently high, but simple estimates suggest that this would
require particle densities of $10^{16}\,{\rm m}^{-3}$ which is orders
of magnitude higher than believed to be the case in lobes or even
hotspots.  Thus it is reasonable to assume that in the vast majority
of cases the observed curvature of frequency spectra arises because of
an intrinsically curved particle spectrum rather than because of
external attenuating influences.

\subsection{The particle spectrum}
\label{sec:particlespectrum}
Curved synchrotron spectra described in
equation\,\ref{eq:curvedspec_lum_nu} suggest an underlying energetic
electron distribution with the shape

\begin{equation}
\ln(N(\gamma))=\ln(N_0)-p\ln(\gamma)-q(\ln(\gamma))^2
\label{eq:curvedspec_N_gamma}
\end{equation}
where $N(\gamma)$ is the differential number of electrons per unit
volume with Lorentz factor $\gamma$ and $N_0$ is the amplitude of that
spectrum extrapolated down to $\gamma=1$. The curvature of the
spectrum is contained in the {\it curvature index} $q$, while the
traditional power-law limit, with index $p$, is recovered when $q=0$. 
We note that when $q>0$, giving a local maximum to the spectrum, that 
the index $p$  (which is the slope of the distribution extrapolated to $\gamma=1$) 
must be negative.    While curved
spectra can also contain low and high energy cutoffs, these are less
critical than in the pure power-law case where the energy density is
dominated by one of these cutoffs which are invariably outside the
observational window.  In the curved case the energy density, total
luminosity and other physically important quantities are determined by
the local maximum in the energy spectrum at $\gpeak$. The
synchrotron energy-loss rate for an electron with Lorentz factor
$\gamma$ in a magnetic field $B$ is given by
\begin{equation}
\dot{E}=-{4\over 3}\sigma_Tc\gamma^2{B^2\over 2\mu_0}=-b_0\gamma^2B^2
\label{eq:edot}
\end{equation}
where $\sigma_T$ is the Thomson cross section and $b_0=1.0588\times
10^{-14}\,{\rm W{Hz}^{-1}T^{-2}}$. While a single particle will
radiate across a range of frequencies, we use here the well-known
approximation that all of the emission is radiated at the single
frequency given by:
\begin{equation}
\nu=\gamma^2{eB\over 2\pi m_e}=b_1\gamma^2 B
\label{eq:gamma_nu}
\end{equation}
with $b_1 =2.799\times 10^{10}\,{\rm Hz\,T^{-1}}$. The differential
energy emitted per unit volume at frequency $\nu$ by a spectrum of energetic particles
with differential energy spectrum per unit volume $N(E)$ is then given by

\begin{equation}
L_\nu d\nu = -\dot{E} N(E)dE = -\dot{E}N(\gamma)d\gamma
\label{eq:curvedspec_lum_edot_N}
\end{equation}
so that 

\begin{equation}
L_\nu={b_0\over 2b_1}B\left({\nu\over b_1 B}\right)^{1/2}N\left(\left({\nu\over b_1 B}\right)^{1/2}\right).
\end{equation}
Inserting the quadratically-curved particle spectrum of
equation\,\ref{eq:curvedspec_N_gamma}, we obtain the required
synchrotron spectrum in the form

\begin{equation}
\ln(L_\nu)=a_0-\left({p-1\over 2}\right)\ln(\hat{\nu})-{q\over 4}\left(\ln(\hat{\nu})\right)^2
\label{eq:curvedspec_lum_nuhat}
\end{equation}
with $\hat{\nu}=\nu/b_1 B$ and where the constant $a_0$ depends on the
magnetic field strength.  We can now relate
the coefficients of the synchrotron spectrum in
equation\,\ref{eq:curvedspec_lum_nu} to those of the particle
spectrum: $a_1=-(p-1)/2$ and $a_2=-q/4$.

A more useful expression for the curved particle distribution replaces
the parameters $N_0$ and $p$ in favour of the position,
$\gpeak$, and differential number density $\npeak\equiv N(\gpeak)$ 
at the local maximum. The energy-dependent slope of the distribution is
\begin{equation}
{d\ln N\over d\ln\gamma}=-p-2q\ln(\gamma)
\label{eq:logN_loggamma}
\end{equation}
so that the differential number of particles has a maximum value at 

\begin{equation}
\gpeak=\exp\left(-{p\over 2q}\right). 
\label{eq:gamma-peak}
\end{equation}
Since $q>0$ the parameter $p$ must be negative for the spectra
characteristically exhibited by radio galaxies and quasars so that a
spectrum with a local maximum at $\gpeak>1$ must be rising at the
point where $\gamma=1$.  The electron distribution can be written in
terms of the position and height of the spectral peak and the
curvature index:
\begin{equation}
\ln(N)=\ln(\npeak)-q\left(\ln(\gamma) - \ln(\gpeak)\right)^2
\label{eq:Npeak_gamma}
\end{equation}
which we can more easily relate to the frequency and intensity of the
maximum synchrotron luminosity.   The particle spectrum is then 
\begin{equation}
N(\gamma)=N_{\rm peak}\exp\left[-q\left(\ln{\left(\gamma\over\gamma_{\rm peak}\right)}\right)^2\right].
\label{eq:particlespectrum}
\end{equation} 
Power-law spectra in radio lobes have traditionally been characterised by a
lower cutoff, $\gamma_{\rm min}$, the value of the spectrum at that
point, $N_{\rm min}$, and the index $p$.
In the curved case we also have three parameters; the position of the
peak, $\gpeak$, the value of the distribution at that Lorentz factor,
$\npeak$, and the curvature index $q$.  The number and energy densities
of the curved distribution are: \eqb n=\int N(\gamma)\,
d\gamma\;\;\;\;{\rm and}\;\;\;\; U_{\rm P}=\int \gamma m_e c^2
N(\gamma)\,d\gamma
\label{eq:defn_defe}
\eqe
which will depend on $\gpeak$, $\npeak$ and $q$, with a very weak
dependence on any low-energy cutoff.  Calculating moments of a
quadratic distribution (on the {\it log-log} plane) is
non-trivial. However, it can be shown (see Appendix) that the
spectrum described in equation\,\ref{eq:particlespectrum}
also has the shape of a {\it log-normal} distribution, which
is the probability distribution for the exponent of a random variable
with a Gaussian distribution. The moments of {\it log-normal}
distributions are well known and allow us to derive expressions for
the number density and energy density in terms
of peak properties ($\gpeak$ and $\npeak$) and spectral curvature
($q$), as follows:
\begin{equation}
n=\sqrt{\pi\over q}e^{1/4q}\gpeak\npeak
\label{eq:defn_defe_defNpeak}
\end{equation}
and
\begin{equation}
U_{\rm P}=\sqrt{\pi\over q}e^{1/q}\gpeak^2 m_{\rm e} c^2 \npeak=
n m_{\rm e}c^2 e^{3/4q}\gpeak.
\end{equation}
Frequency-resolved observations of a synchrotron spectrum will allow
us to calculate $q$ directly but, without knowledge of the magnetic
field, we cannot directly infer values for $\gpeak$ and $\npeak$
unambiguously from the peak emission. Relating the peak of the
energy spectrum to the frequency, and intensity, of the maximum
emission depends not only on $B$ but also on the spectral curvature as
we now show.

\subsection{The synchrotron spectrum}

\begin{figure*}
\includegraphics[width=\textwidth]{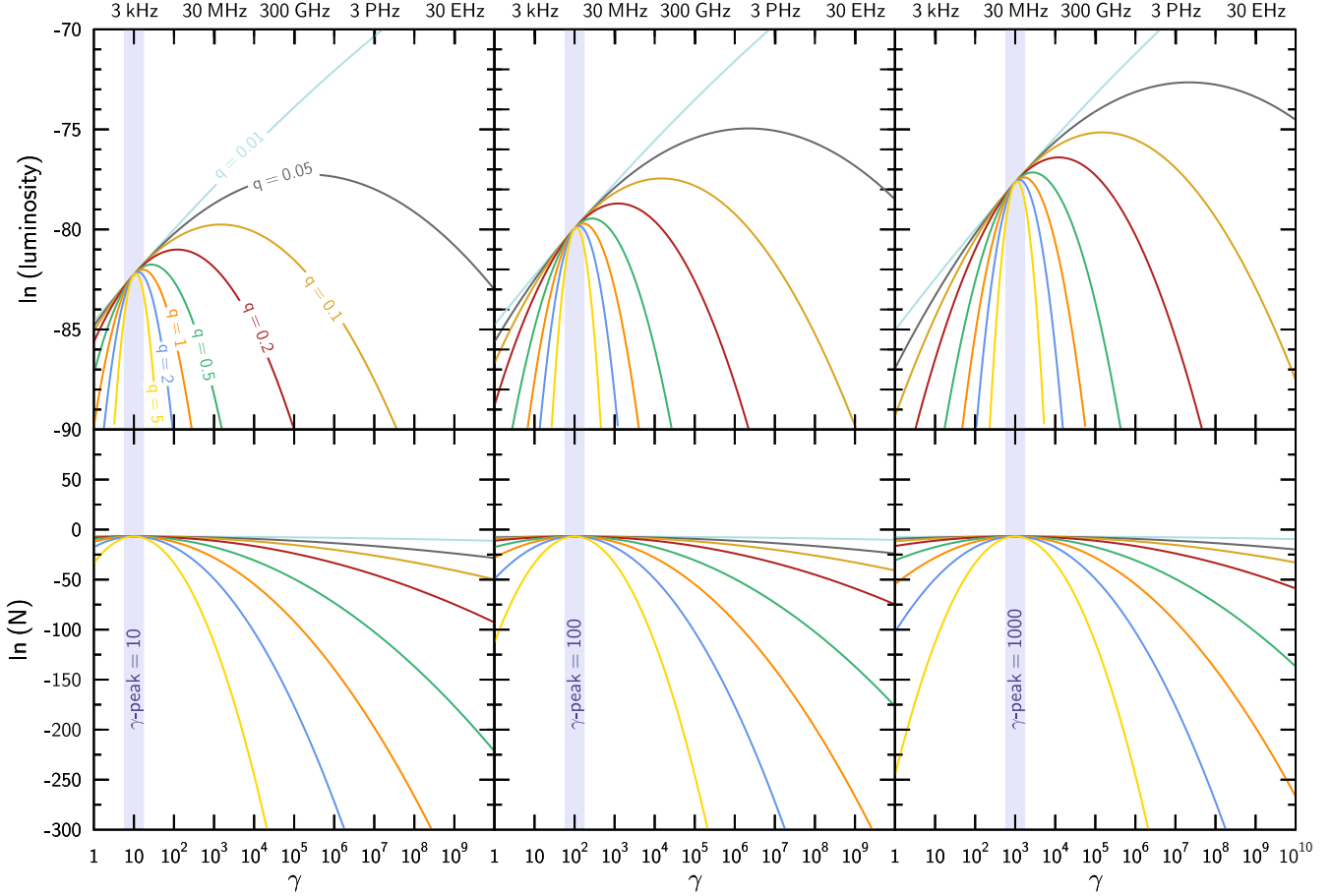}
\caption{\label{fig:peaksmove} This figure illustrates how for
  different values of $q$ the energies of the electrons at the peak of
  the particle energy distribution are distinctly different from those
  electrons that give rise to the most intense synchrotron emission in
  the frequency spectrum.  }
\end{figure*}

The underlying electron distribution will, in the presence of a
magnetic field, give rise to synchrotron emission with a local maximum
at frequency $\numax$ where the differential number density will
determine the value of the emissivity at this frequency which we term
$\lmax$. Since the curvature of the synchrotron spectrum is $q/4$ it can 
be shown, in an analogous way to equation\,\ref{eq:Npeak_gamma}, 
that the frequency spectrum can be written in the form:
\begin{equation}
\ln(L_\nu)=\ln(\lmax)-{q\over 4}\left(\ln(\nu)-\ln(\numax)\right)^2.
\label{eq:Lnu}
\end{equation}
It is tempting to assume that electrons with a Lorentz factor of
$\gpeak$ give rise to the emission at $\numax$, but this is not the
case as Fig\,\ref{fig:peaksmove} illustrates. The emissivity of a
single particle scales with $\gamma^2$ so that particles which are
slightly more energetic than those at $\gpeak$, although fewer in
number, will radiate more intensely. This is most easily seen by
returning to the approximation where an electron with a Lorentz factor
$\gamma$ emits at a single frequency $\nu\propto\gamma^2$ so that

\eqb
L_\nu\propto \dot{E}(\gamma)N(\gamma){d\gamma\over d\nu}\propto \gamma^2 N(\gamma)\nu^{-1/2}\propto \gamma N(\gamma),
\label{eq:Lnupropto}
\eqe 
and the emission is at its most intense not for the Lorentz
factor that maximises $N(\gamma)$ (i.e. $\gpeak$) but rather for the
value that maximises $\gamma N(\gamma)$ which is found from

\begin{equation}
{d\ln (\gamma N)\over d\ln\gamma}=1-p-2q\ln(\gamma).
\label{eq:dlngammaN}
\end{equation}
Thus, the particles responsible for the maximum emission in the
frequency spectrum are shifted upwards in energy from the particles at
the peak of the energy distribution, and have a Lorentz factor given
by

\begin{equation}
\gamma(\numax)=\gpeak e^{1/2q}.
\label{eq:gammanumax}
\end{equation}
Figure\,\ref{fig:peaksmove} illustrates the effect of curvature for a range of values 
between $q=0.01$ and $q=5$. For example, when $q= 0.2$ there is more
than an order of magnitude difference between the energies of
electrons at the peak of distribution and those electrons that give
rise to the most intense synchrotron emission.   We can now relate
$\numax$ and $\lmax$ to the corresponding parameters for the electron
distribution, $\gpeak$ and $\npeak$ as follows:

\begin{equation}
\numax = b_1\gamma(\numax)^2 B=b_1 e^{1/q}\gpeak^2 B
\end{equation}
\begin{eqnarray}
\lmax &=& {b_0\over 2b_1}B\gamma(\numax) N(\gamma(\numax)) \nonumber \\
          &=& {b_0\over  2b_1}e^{1/4q}B\gpeak {\npeak}. 
\label{eq:numax_to_gpeak}
\end{eqnarray}
\newpage
\onecolumn

\subsection{Equipartition magnetic field strength}

We can express $\gpeak$, $\npeak$, $n$ (measured in ${\rm m}^{-3}$) and $U_{\rm P}$ 
(${\rm J/m}^3$) in terms of the observable quantities $\numax$, $\lmax$ and $q$, subject to the 
uncertainty in the magnetic field
\eqb
\gpeak&=&{e^{-1/2q}\over \sqrt{b_1}}\numax^{1/2}B^{-1/2}=
5.98\times 10^3e^{-1/2q}
\left({\numax\over 1\,{\rm GHz}}\right)^{1/2}
\left({B\over 1\,{\rm nT}}\right)^{-1/2}
\label{eq:gpeaklum}\\
\npeak&=&{2b_1^{3/2}\over b_0}e^{1/4q}\lmax\numax^{-1/2}B^{-1/2}
=8.85\times 10^{-6}e^{1/4q}
\left({\lmax\over 10^{-35}\,{\rm W\,m^{-3}\,Hz^{-1}}}\right)
\left({\numax\over 1\,{\rm GHz}}\right)^{-1/2}
\left({B\over 1\,{\rm nT}}\right)^{-1/2}
\label{eq:npeaklum}\\
n&=&{2b_1\over b_0}\sqrt{\pi\over q}\lmax B^{-1}
=5.29\times 10^{-2}\sqrt{\pi\over q}
\left({\lmax\over 10^{-35}\,{\rm W\,m^{-3}\,Hz^{-1}}}\right)
\left({B\over 1\,{\rm nT}}\right)^{-1}
\label{eq:nlum}\\
U_{\rm P}&=&{2\sqrt{b_1}\over b_0}m_ec^2
\sqrt{\pi\over q}e^{1/4q}\lmax\numax^{1/2}B^{-3/2}=
2.59\times 10^{-11}\sqrt{\pi\over q}e^{1/4q}
\left({\lmax\over 10^{-35}\,{\rm W\,m^{-3}\,Hz^{-1}}}\right)
\left({\numax\over 1\,{\rm GHz}}\right)^{1/2}
\left({B\over 1\,{\rm nT}}\right)^{-3/2}
\label{eq:elum}
\eqe
The ratio of particle to magnetic energy density is then
\eqb
\theta\equiv {U_{\rm P}\over U_{\rm B}}=65.08
\sqrt{\pi\over q}e^{1/4q}
\left({\lmax\over 10^{-35}\,{\rm W\,m^{-3}\,Hz^{-1}}}\right)
\left({\numax\over 1\,{\rm GHz}}\right)^{1/2}
\left({B\over 1\,{\rm nT}}\right)^{-7/2}
\label{eq:thetaequip}
\eqe
The magnetic field strength in equipartition with the energetic electrons is then (with $q_{0.2}\equiv q/0.2$)

\eqb
\bequip= 2.01\left({e^{1/2q}\over q}\numax \lmax^2\right)^{1/7}
= 6.98\,
\left({e^{1/2q_{0.2}}\over q_{0.2}}\right)^{1/7}
\left({\numax\over 1\,{\rm GHz}}\right)^{1/7}\left({\lmax\over 10^{-35}\,{\rm W\,m^{-3}\,Hz^{-1}}}\right)^{2/7}\,{(\rm nT})
\label{eq:Bequip}
\eqe
\twocolumn
\noindent where $B_{\rm equip}$ is here expressed in nano-Tesla.
Therefore if we can be confident that the magnetic field energy of a plasma is
in equipartition with its particle energy then we have obtained
$B$-field measuring machinery in terms of the observables $\nu_{\rm
  max}$, $\lmax$ and $q$.  
  
The magnetic field that minimises the total energy, $B_{\rm min}$, can be obtained 
from the expression for the total electron and magnetic field energy densities

\eqb
U=U_{\rm P}+U_{\rm B}={c_1\over B^{3/2}}+c_2 B^2
\eqen
where $c_1$ and $c_2$ are constants such that $B_{\rm equip}=(c_1/c_2)^{2/7}$. 
The energy density is a minimum for $B_{\rm min}=(3c_1/4c_2)^{2/7}=0.92 B_{\rm equip}$ 
so that, as in the power law case, the equipartition and minimum energy $B$ fields are 
comparable.

\subsection{Cygnus A}

\begin{figure}
  \centering
\vbox{
  \includegraphics[width=0.95\columnwidth]{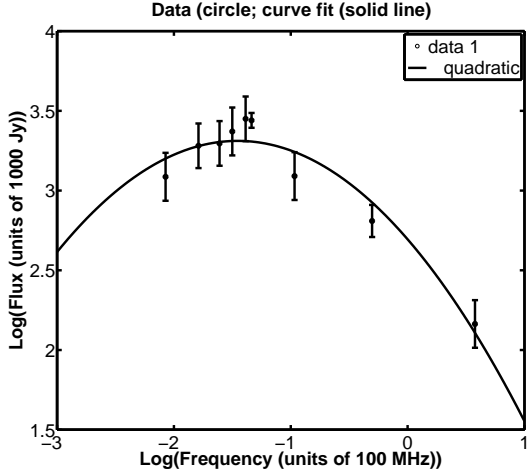}
  }
  \caption{ \label{fig:cyga} Plot of the radio frequency spectrum of
    the prototypical powerful classical double radio galaxy Cygnus\,A
    and the best quadratic fit to these points.  }
\end{figure}

We now apply these results to the lobes of the prototypical classical
double radio galaxy, Cygnus\,A.  In Figure\,\ref{fig:cyga} we show both
the observed spectra and best-fit quadratic, on the {\it log-log},
plane. The equation for the best-fit curve is

\eqb
y=-0.29x^2-0.85x+2.7
\eqe
where

\begin{equation}
y\equiv\ln\left({F_{\nu}\over 10^3\,{\rm Jy}}\right)\;\;\;\;{\rm and}\;\;\;\;
x\equiv\ln\left(\nu\over 100\,{\rm MHz}\right).
\end{equation}
The relation between emissivity and frequency determines $q$
through the quadratic term,

\begin{equation}
\ln(L_{\nu})=-{q\over 4}(\ln\nu)^2+...
\end{equation}
Converting from flux density ($F_\nu$) to emissivity ($L_\nu$)  (which involves the
distance to the source and its volume) and normalising the frequency
to arbitrary units will not change the coefficient of the quadratic
term.  Therefore, for Cygnus A we have

\eqb
{q\over 4}=0.29\;\;\;\Rightarrow\;\;\; q=1.16
\eqe
and the emission has a maximum at 

\eqb \nu_{\rm max}=23\,{\rm MHz}\;\;\;\;{\rm with}\;\;\;\; F_{\nu_{\rm
    max}}=2.77\times 10^4\,{\rm Jy}.  \eqe Since the source is at a
distance of $D=600\,{\rm Mlyr}$ and has a volume estimated to be
$V\approx 1.73\times 10^{63}\,{\rm m^3}$, the peak emissivity is
\eqb
L_{\nu_{\rm max}}=6.52\times 10^{-35}\,{\rm Wm^{-3}Hz^{-1}}.
\eqe
Using equations \ref{eq:gpeaklum}, \ref{eq:npeaklum}, \ref{eq:nlum} and 
\ref{eq:elum} we can express $\gpeak$, $\npeak$, $n$ and $e$ for Cygnus\,A in
terms of the magnetic field $B_{\rm nT}$ (in units of $1\,{\rm nT}$),
\eqb
\gpeak&=&5.91\times 10^2 B_{\rm nT}^{-1/2}\\
\npeak&=&4.71\times 10^{-4}B_{\rm nT}^{-1/2}\\
n&=&0.57 B_{\rm nT}^{-1}\\
U_{\rm P}&=&5.21\times 10^{-11}B_{\rm nT}^{-3/2}.
\eqe
The equipartition field is then $B_{\rm nT}=4$ implying a peak to
the electron distribution at $\gpeak=296$, a number density of
$n=0.14\,{\rm m^{-3}}$ and an energy density of $U_{\rm P}=6.37\times
10^{-12}\,{\rm Jm^{-3}}\approx 40\,{\rm eV\,cm^{-3}}$.

However, there is compelling evidence from multiwavelength
observations of Cygnus A \citep{Steenbrugge2008} that a peak in the
electron distribution along the jet must be above a few times $10^3$ in Lorentz
factor. Taking $\gpeak=10^4$ we can invert the above argument to give
$B_{\rm nT}\approx 3.49\times 10^{-3}$, which is well below
equipartition.   In this case we have a weak field, peaked at a high Lorentz factor, and 
with a high number density of energetic electrons, $n\approx160\,{\rm m^{-3}}$. 
The existence of a high number density of particles at $\gpeak=10^4$ 
has important implications for the relativistic Sunyaev-Zeldovich effect from this 
source, as will be shown in the next section.

\section{Curved electron spectra and the CMB}

In addition to their acceleration in the background magnetic field the
relativistic electrons will also interact with the CMB. Two
interesting, and potentially observable, effects can then
arise. First, Inverse Compton scattering off the CMB (ICCMB) will
result in X-ray emission (for $\gpeak\sim 10^3$). This spectrum will
be peaked at a frequency determined by $\gpeak$, $q$ and the frequency
at which the maximum intensity occurs in the CMB spectrum. Second,
photons in the Rayleigh-Jeans regime of the CMB will be upscattered by
the energetic electrons leading to an observed reduction in the CMB
temperature: the Sunyaev Zeldovitch (SZ) effect. We examine each of
these below.

An energetic electron in a photon field with energy density $U_{\rm
  rad}$ will lose energy through Inverse-Compton scattering at the
rate,

\begin{equation}
\dot{E}=-{4\over 3}\sigma_Tc\gamma^2U_{\rm rad}.
\label{eq:edoticcmb}
\end{equation}
The electrons scatter the
ambient CMB photons producing higher energy photons (of frequency
$\nu_{\rm X}$) according to:
\begin{equation}
\frac{\nu_{\rm X}}{\nu_{\rm CMB}} = \displaystyle \left(  \frac{4}{3} \right)\gamma^2 - \frac{1}{3}\approx {4\over 3}\gamma^2.
\label{eq:CMB_ratio}
\end{equation}
As in the synchrotron case, the maximum intensity X-ray emission will
not correspond to ICCMB scattering off $\gpeak$ particles but rather
particles with a Lorentz factor of $\gamma(\numax)=\gpeak e^{1/2q}$;
i.e. the maximum intensities in both the radio synchrotron and ICCMB
emission are caused by particles of the same Lorentz factor (the one
for which $\gamma N(\gamma)$ is a maximum).  This occurs because the
dependence of $\dot{E}$ and the dominant synchrotron emission frequency have the
same dependence on $\gamma$. Hence, the ratio of the peak emission in
synchrotron to that in X-rays will only depend on the ratio of
magnetic field energy density to the equivalent energy density in the
CMB (as in the power-law case).  Moreover, if we were to observe a
maximum in the ICCMB spectrum at $\nu_{\rm IC,max}$ then we can relate
that to the shape of the particle distribution (for $q\ne0$) through
\begin{equation}
\nu_{\rm IC,max}\approx {4\over 3} \gpeak^2 e^{1/q}\nu_{\rm CMB}.
\end{equation}
With $\gpeak = 10^3$ and $q=0.2$, a source at redshift $z$ would then
have an ICCMB spectrum peaked at
\begin{equation}
{h\nu_{\rm IC,max}\over 1\,{\rm keV}}\approx
130(1+z) \left({\gpeak\over 10^3}\right)^2e^{1/q_{0.2}}
\label{eq:iccmb_max}
\end{equation}
in the hard X-ray to soft gamma-ray region.  It is useful to compare
this prediction with that of \citet{Blu06} who consider the ICCMB
spectrum from a giant radio galaxy at redshift $z \sim 2$ where the
electron spectrum is a power law with a low-energy cutoff at
$\gamma_{\rm min}\gg 1$. When the index $p>2$, the differential number
and energy densities are dominated by particles at this low energy
cutoff. The X-ray ICCMB emission will therefore peak at a frequency of
$\nu_{\rm X,min}\approx \gamma_{\rm min}^2\nu_{\rm CMB}$, with the
emission falling off rapidly for lower frequencies and with a power
law above $\nu_{\rm X,min}$. But in the curved case, just as with
synchrotron emission, it is particles with a Lorentz factor of $\gpeak
e^{1/2q}$ that will give rise to the most intense ICCMB
emission. Since the emitted frequency scales quadratically with
$\gamma$ this can have a dramatic effect on $\nu_{\rm IC,max}$. For
$\gpeak=10^3$ and $q=0.2$ this pushes the peak emission into the hard
X-ray to soft gamma-ray region. For a low-energy power-law cutoff at
$\gamma_{\rm min}=10^3$ (and with $p>2$) the ICCMB emission peaks at
frequencies of about a few keV --- some two orders of magnitude below
a curved spectrum with $\gpeak=10^3$ and $q=0.2$.  While the position
of the peak can be quite different, depending on $q$, for curved and
power-law spectra with $\gpeak=\gamma_{\rm min}$, the intensity of the
emission will, of course, depend in each case on the number density of
energetic particles.

The SZ effect predicts a reduction (for the Rayleigh-Jeans regime) in
the temperature of the CMB for the (optically thin) scattering of
photons through a hot electron gas in a galaxy cluster. When the CMB photons are 
scattered by relativistic, as opposed to thermal, particles the temperature is also reduced 
\citep{Bir99,McKinn91} and is related to the optical depth, $\tau$, by

\eqb
{\Delta T\over T} \approx - \tau
\label{eq:deltaToverT}
\eqe
where 
\eqb
\tau = 2\sigma_T n R_{\rm lobe},
\label{eq:opticaldepth}
\eqe
where $R_{\rm lobe}$ is the path length through the plasma lobe. 
Our number density of target
electrons is no longer a thermal distribution but non-thermal and
curved so that
\eqb
n=\sqrt{\pi\over q}e^{1/4q}\gpeak \npeak.
\eqe
We would therefore expect a reduction of the CMB temperature in the
direction of a lobe of the order
\eqb
{\Delta T\over T}\approx -2 \sigma_T n R_{\rm lobe}\approx -2\sqrt{\pi\over q}e^{1/4q} 
\gpeak \npeak \sigma_T R_{\rm lobe}.
\label{eq:newDeltaToverT}
\eqe
From equations \ref{eq:gpeaklum} and \ref{eq:npeaklum}, $\gpeak\npeak$ depends on the 
peak synchrotron emissivity, magnetic field strength and index $q$
\eqb
\gpeak\npeak = 5.29\times 10^{-2}e^{-1/4q}
\left({\lmax\over 10^{-35}\,{\rm W\,m^{-3}\,Hz^{-1}}}\right)
\left({B\over 1\,{\rm nT}}\right)^{-1}.
\label{eq:gpeaknpeak}
\eqe
The relativistic SZ effect gives a decrease in temperature that can be related to the 
observable peak and curvature of the synchrotron spectrum, and the magnetic field strength. 
For a lobe with a scale $R=10\,{\rm kpc}$ we get
\eqb
{\Delta T\over T}\approx {3.85\times 10^{-9}\over\sqrt{q}}
\left({R_{\rm lobe}\over 10\,{\rm kpc}}\right)
\left({\lmax\over 10^{-35}\,{\rm W\,m^{-3}\,Hz^{-1}}}\right)
\left({B\over 1\,{\rm nT}}\right)^{-1}
\label{eq:newDeltaToverTnumbers}
\eqe 
giving a temperature reduction of a few nK for Cygnus A if the magnetic field 
is in equipartition and of the order of several nT.

However, in the previous section it was argued that Cygnus A 
might contain a high density of particles at $\gpeak=10^4$ in order to 
explain the synchrotron emission in a magnetic field that is well below 
equipartition, $B_{\rm nT}=3.46\times 10^{-4}$. A high density of ultrarelativistic 
particles in a such a weak field gives a fractional reduction in the CMB 
temperature of 
\eqb
{\Delta T\over T}\approx 6.69\times 10^{-5} 
\left({R_{\rm lobe}\over 10\,{\rm kpc}}\right).
\label{eq:DeltaToverTnumbersCyga}
\eqe
which is at a level that is detectable by current technology. 

\section{Summary}
\label{sec:summary}
We have presented a formalism which gives the easy derivation of
important physical quantities of plasma lobes (energy, magnetic field
strength, optical depth hence fraction of CMB upscattered) from the
simple characterisation of their spectra as quadratic on the
log-frequency log-luminosity plane.  We also point out that the peak
of the curved distribution of particles' Lorentz factors in a plasma
lobe will not map directly to the frequency at which the peak
intensity is observed --- this is true for both the synchrotron emission, and the
ICCMB scattered emission observed in the X-rays or higher energies.  We have also
quantified the contribution to CMB anisotropies from such relativistic
plasma lobes as a function of the extent to which such plasma lobes
exhibit equipartition of energies stored in particles and magnetic
fields.

\section*{Acknowledgments}

The authors are grateful for the support of Science Foundation
Ireland, the Research Centre of St John's College Oxford, the Royal
Irish Academy and the Royal Society.  KMB thanks the Royal Society for
a University Research Fellowship. PD acknowledges support from Science 
Foundation Ireland under grant 05/RFP/PHY0055.

\bibliographystyle{mn2e} 

\section{Appendix}

The intrinsically-curved energy spectrum can be re-cast in terms of a
log-normal distribution

\begin{eqnarray}
N(\gamma) &=& N_{\rm
  peak}\exp\left[-q\left(\ln{\left(\gamma\over\gamma_{\rm
          peak}\right)}\right)^2\right] \nonumber \\
&=& {{\hat n}\over \sigma \sqrt{2\pi}}{1\over\gamma}\exp\left[-{(\ln\gamma - \mu)^2\over 2\sigma^2}\right]
\label{eq:lognormal}
\end{eqnarray}
using the substitutions,
\eqb
\sigma&=&{1\over\sqrt{2q}}\nonumber\\
\mu&=&{1\over 2q}+\ln\gpeak\nonumber\\
{\hat n}&=&\sqrt{\pi\over q}e^{1/4q}\gpeak N_{\rm peak}.
\eqe
In probability theory if a random variable $y$ has a Gaussian
probability distribution, with mean $\mu$ and variance $\sigma^2$,
then its exponent, $x=e^y$,  
has a log-normal distribution \citep{Aitch57} . With probability densities $P(x)$ and $P(y)$ 
this can be seen from the change of variable, 
\eqb
P(x)=P(y(x))\left|{dy\over dx}\right|={1\over \sigma\sqrt{2\pi}}{1\over x}\exp\left[-{(\ln x - \mu)^2\over 2\sigma^2}\right]
\eqe
where $y$ takes on all values between $-\infty$ and $\infty$ while $x$
lies between $0$ and $\infty$. That the energetic particle
distribution in giant radio galaxy lobes 
has a shape close to the exponential of a normally-distributed
variable is clearly saying something important about the physics
involved in the  
acceleration process. For
our current purposes we can exploit results on the first three moments
of a log-normal function  
to derive useful approximations for the particle number density,
energy density and total synchrotron power. The expectation value of
$x^m$ is given by 
\eqb
E[x^m]\equiv \int_0^\infty x^m P(x)\,dx=e^{m\mu +n^2\sigma^2/2}.
\eqe
We cannot apply this exact result to the log-normal form of the
particle distribution function since $\gamma \ge 1$. However, we are
interested in cases where on observational grounds, $\gpeak\gg 1$ and
the number density is dominated by particles near that peak
energy. For this reason, and also since we don't know a priori where any low
energy cut-off might occur, the number density can be approximated by 
the zeroth moment ($m=0$) of a log-normal distribution,
\eqb
n={{\hat n}\over \sigma \sqrt{2\pi}}\int_1^\infty {1\over\gamma}\exp\left[-{(\ln\gamma - \mu)^2\over 2\sigma^2}\right]\,d\gamma\approx {\hat n}E[\gamma^0]={\hat n}.
\eqe
The number density is therefore related to the peak properties and spectral curvature through
\eqb
n=\sqrt{\pi\over q}e^{1/4q}\gpeak N_{\rm peak}.
\label{eq:numberdensity}
\eqe
For $m=1$ and $m=2$ we have
\eqb
E[\gamma]=e^{3/4q}\gpeak\;\;\;{\rm and}\;\;\;
E[\gamma^2]=e^{2/q}\gpeak^2.
\eqe
The particle energy density is then
\eqb
U_{\rm P}=nm_e c^2 E[\gamma]=\sqrt{\pi\over q}e^{1/q} \gpeak^2 m_e c^2N_{\rm peak},
\eqe
and the second moment gives the total synchrotron luminosity in terms
of peak properties and curvature, 
\begin{eqnarray}
P &=& {4\over 3}\sigma_T cVU_{\rm B} nE[\gamma^2] \nonumber \\
&=& {4\over 3}\sqrt{\pi\over q} e^{9/4q}\sigma_T cVU_{\rm B} \gpeak^3 N_{\rm peak}.
\label{eq:totalsynclumpeaks}
\end{eqnarray}
We make use of these relationships in
Section\,\ref{sec:particlespectrum}.  


\end{document}